\documentclass[prx,reprint,noshowpacs,superscriptaddress,floatfix,letterpaper,notitlepage]{revtex4-2}
\usepackage{amsmath,amssymb,amsbsy,amsfonts,amsthm,bbm,bm,mathtools,mathrsfs}
\usepackage{color}
\usepackage{physics}
\usepackage[dvipsnames]{xcolor}
\pagecolor{white}
\usepackage[colorlinks=true,citecolor=Cerulean,linkcolor=RubineRed,urlcolor=Cerulean]{hyperref}
\usepackage{empheq}
\usepackage{stackengine}
\usepackage{yfonts}
\usepackage[normalem]{ulem}
\usepackage{enumitem}
\usepackage{subfiles}
\usepackage{sidecap}
\sidecaptionvpos{figure}{c}

\DeclareMathOperator*{\argmin}{arg\,min}

\DeclareMathAlphabet{\mathpzc}{OT1}{pzc}{m}{it}
\newcommand{\intercept}{a}
\newcommand{\ointercept}{\hat a}
\newcommand{\slope}{\boldsymbol{b}}
\newcommand{\oslope}{\boldsymbol{\hat b}}
\newcommand{\soslope}{\hat b}

\newcommand{\speedtime}{\tau_{\!\scriptscriptstyle\mathcal{X}}}
\newcommand{\chem}{{\!\scriptscriptstyle\textrm{chem}}}

\begin{document}

\title{Thermodynamic speed limits from the regression of information}

\author{Schuyler~B.~Nicholson}
\affiliation{Department of Chemistry,\
 University of Massachusetts Boston,\
 Boston, MA 02125, USA
}
\affiliation{Department of Chemistry,\
  Northwestern University,\
  Evanston, IL 60208
}
\author{Jason~R.~Green}
\email[]{jason.green@umb.edu}
\affiliation{Department of Chemistry,\
  University of Massachusetts Boston,\
  Boston, MA 02125, USA
}
\affiliation{Department of Physics,\
  University of Massachusetts Boston,\
  Boston, MA 02125, USA
}

\date{\today}

\begin{abstract}

Irreversible processes accomplished in a fixed time involve nonlinearly coupled flows of matter, energy, and information.
Here, using entropy production as an example, we show how thermodynamic uncertainty relations and speed limits on these nonlinear processes derive from linear regression.
These uncertainty relations hold for both passive and actively-driven nonequilibrium processes and all have a mathematical form that mirrors uncertainty relations in quantum mechanics.
Using optimal linear models, we show that information-theoretic variables naturally give physical predictions of the equation of motion on statistical manifolds in terms of physical observables.
In these models, optimal intercepts are related to nonequilibrium analogs of Massieu functions/thermodynamic potentials, and optimal slopes are related to speed limits on collections of thermodynamic observables.
Within this formalism, the second law of thermodynamics has a geometric interpretation as the nonnegativity of the slope and constrains the equation of motion.
Overall, our results suggest that unknown relationships between nonequilibrium variables can be learned through statistical-mechanical inference.

\end{abstract}

\maketitle

\medskip

Any process taking a physical system on an excursion from equilibrium will incur a thermodynamic cost: energy dissipated as heat, wasted free energy, or the production of entropy. Optimizing thermodynamic benefits such as work and predicting bounds on efficiency, requires relationships between variables that have been hypothesized, tested, and validated against experiments~\cite{callen85}.
For example, we predict spontaneous chemical and physical changes through the fundamental relations between thermodynamic potentials and a set of displacements in experimentally controllable parameters, weighted by their conjugate thermodynamic driving forces.
A closed system might be subject to controlled, reversible variations in volume $V$ and internal energy $U$ modulated by the forces of $1/T$ and $-P/T$.
Predictions of stability and spontaneity are made with the appropriate thermodynamic potential, the entropy $S(U,V)$, through the linear relationship $dS=dU/T-PdV/T$.
A principal success of thermodynamics is the ability to predict the changes in thermodynamic potentials that accompany intentional manipulations of controllable parameters.
This task, however, is more challenging for systems that are small, subject to fluctuations, and driven strongly out of equilibrium. We are often forced into assumptions about the dynamics, the nature of the fluctuations, the driving protocol, the noise, or the ``distance'' from equilibrium~\cite{jarzynski11,seifert2012stochastic,marsland_2017}.

Like thermodynamics, \textit{regression} involves the investigation of quantitative, predictive relationships between variables.
Regression dates back to Galton~\cite{galton1886regression} who used the term and associated methods for the evolution of genetic traits and the statistical tendency for large deviations to subsequently regress to the mean.
In statistical physics, Onsager~\cite{onsager1931}
used this term in a similar sense in his regression hypothesis for nonequilibrium fluctuations.
Currently, though, with the advancements in machine learning, regression is often used to indicate a measure of the statistical relationship between random variables.
It is now an essential part of supervised methods~\cite{friedman2001elements}, which have recently been used to infer time's arrow~\cite{seif_machine_2021}.
An open question is whether statistical prediction through regression might already be embedded to some extent in the statistical mechanics of nonequilibrium processes.

We set out to answer this question from the perspective of thermodynamic speed limits and uncertainty relations, which have recently set bounds on nonequilibrium fluctuations with the entropy production~\cite{horowitz2019thermodynamic}.
Uncertainty relations have traditionally belonged to quantum mechanics,
with Heisenberg's famous uncertainty principle being a prime example~\cite{Messiah61}. The time-energy uncertainty relation can be also be cast as a quantum speed limit on the evolution of quantum systems between distinguishable states~\cite{mandelstam1945uncertainty,margolus1998maximum,delcampo13,DeffnerLutz13,Luispe19,Zych06}. Classical uncertainty relations, known since the $1950$'s~\cite{mandelbrot1956outline,uffink1999thermodynamic,schlogl1988thermodynamic},
have received renewed interest due to the thermodynamic
uncertainty relation~\cite{pietzonka2017finite,maes2017frenetic},
proposed~\cite{barato2015thermodynamic} and proven through large deviation
theory~\cite{horowitz2017proof}. 
Rapid progress has extended early steady-state results to other non-equilibrium regimes~\cite{gingrich2016dissipation,van2020unified}, underdamped Langevin dynamics~\cite{van2019uncertainty}, information geometry~\cite{falasco2020dissipation,dechant2020fluctuation,nicholson2020time,nicholson2018nonequilibrium} and the Cram{\'e}r-Rao inequalities~\cite{hasegawa2019uncertainty,ItoDechant2020}, and multiple dimensions~\cite{guarnieri2019thermodynamics,Dechant2020Correlations}.

\begin{SCfigure*}[][t]
\centering
\resizebox{0.6\textwidth}{!}{%
\begin{minipage}[l]{\textwidth}
\includegraphics[width=0.99\textwidth,angle=0,clip]{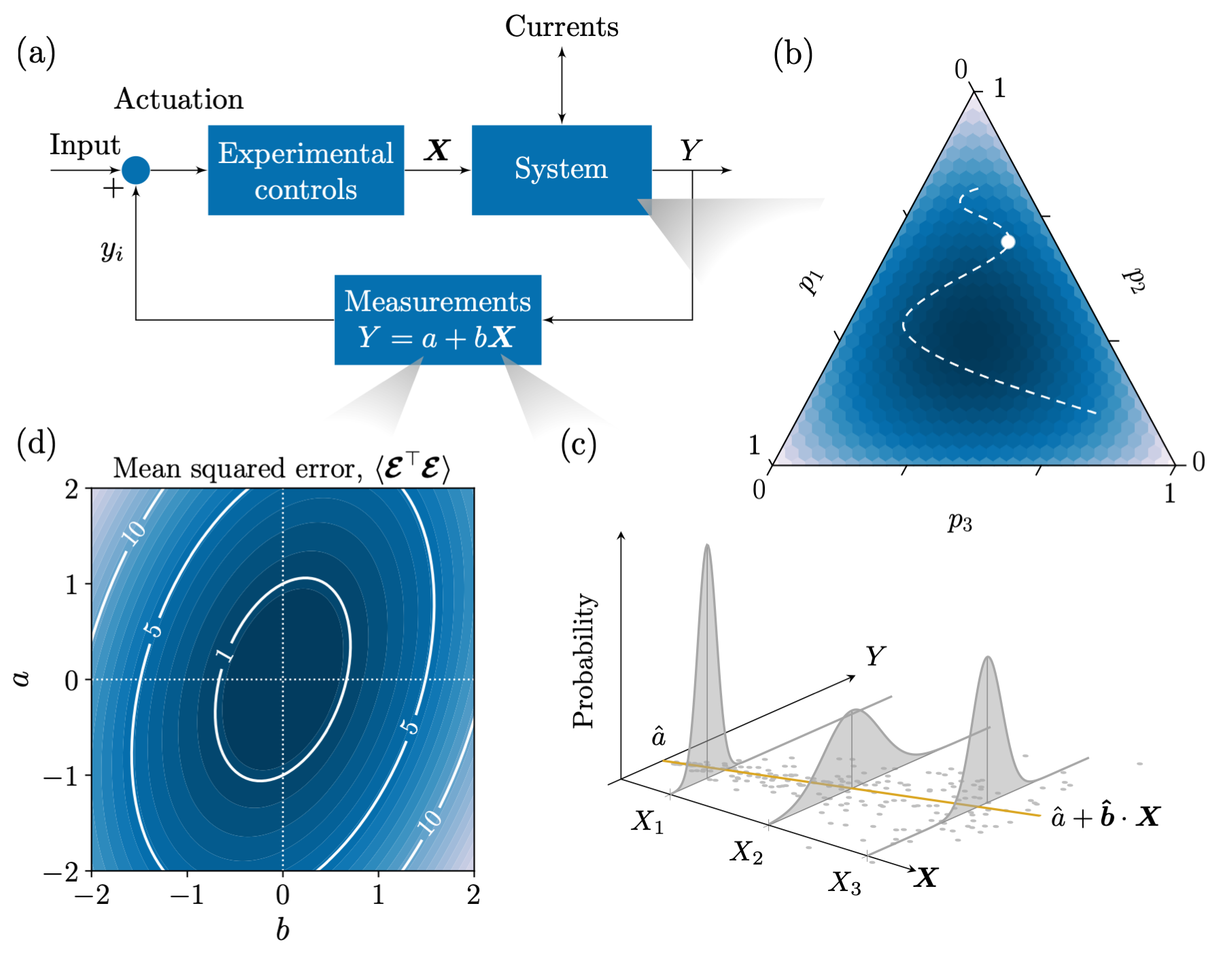}
\end{minipage}}
\caption{\label{fig:fig1}\footnotesize{(a) Schematic illustrating the statistical prediction of relationships between nonequilibrium observables. Driven away from equilibrium, a set of variables $\boldsymbol{X}$ are measurable or controlled as the system is subject to nonequilibrium currents of matter, energy, entropy, and information with the surroundings. (b) As a result, the probability distribution over configurations will evolve in time over a statistical manifold; a probability 2-simplex (made with~\cite{pythonternary}) with the time evolution of a system and the time point chosen for the regression hypothesis. Darker color indicates higher Shannon entropy.
(c) While away from thermodynamic equilibrium, an unknown function of the control variables, $Y$, also evolves in time.
A regression hypothesis models its statistical relationship to control variables $\boldsymbol{X}$, e.g., $Y=a+\slope\cdot\boldsymbol{X}$.
For example, the linear regression of $-\dot I=d\ln p/dt$ on $X_\omega=I-\beta\epsilon$ has an optimal intercept $\ointercept$ determined by the free entropy, $\beta F$, and the optimal slope, $\soslope$, related to the entropy production rate, $\sigma$, through $\soslope=\sigma/\Delta X_\omega^2$.
(d) Optimal predictions of the relationship are found by minimizing the mean squared error $\langle \boldsymbol{\mathcal{E}}^\top\boldsymbol{\mathcal{E}}\rangle$ as a function of the regression parameters $a$ and $\boldsymbol{b}$.
At the minimum, the optimal parameters are $\ointercept$ and $\oslope$, respectively.
Thermodynamic uncertainty relations and speed limits derive from the mean squared error and the optimal slope.}}
\end{SCfigure*}

Here, by analyzing the regression of thermodynamic variables, we show that even the simplest linear statistical models can give the multidimensional thermodynamic uncertainty relations~\cite{barato2015thermodynamic,horowitz2019thermodynamic} and the time-information uncertainty relation--a classical speed limit on fluxes of matter, energy, and entropy~\cite{nicholson2020time}.
These uncertainty relations directly derive from the fluctuations of the optimal linear predictor for the relationship between variables, suggesting a framework for a wide class of observables independent of the stochastic dynamics away from stationary states.
Our results suggest a framework for statistical-mechanical inference of the functional relationships between observables away from thermodynamic equilibrium. We illustrate this framework using optimal linear regression of natural variables that give speed limits on the rate of entropy production of the system.

\section{Thermodynamic regression}

\subsection{Prediction of statistical relationships}

Consider a physical system with currents of matter and energy within and through its boundaries driven by the manipulation of an external control parameter or the removal of a constraint, Fig.~\ref{fig:fig1}(a).
These currents might converge to constant values, sustained by external reservoirs.
Or, the system might be driven strongly, transiently, and perhaps with
large fluctuations that prevent macroscopic observables from establishing
steady values.
How might we learn an unknown property of the system from a set of measured observables? Here, we translate the experimental design into a statistical inference problem for nonequilibrium processes.
Suppose we want to predict a random variable, $Y$, that is a property of the system. 
However, because of experimental constraints, we only have the ability to measure a set of $N$ predictor variables $\boldsymbol{X}:=(X_1,X_2,\ldots,X_N)^\top$, a vector of random variables.
All quantities here can be time dependent unless explicitly stated otherwise.

Regardless of the nonequilibrium process, we make a minimal assumption that the dynamical evolution smoothly transforms the probability, $p[z,\lambda(t)] = p(z,t) = p(z)$, of each state $z$ at time $t$ with a rate $\dot{p}(z) = dp(z)/dt$, Fig.~\ref{fig:fig1}(b).
The dynamics away from equilibrium are usually difficult to determine experimentally and often studied with an appropriate physical model (e.g., Langevin dynamics, master equations).
Instead, we will take another approach and learn the equation of motion $d\boldsymbol{p}/dt$ from a set of measured observables.
The probability distribution we take to be defined over an ensemble of identical systems.
To quantitatively describe the nonstationary behavior of the system, we model the relationship between the intrinsic rate at which the distribution over configurations changes, $Y\to -\dot{I}:=d_t\ln p(t)$, and a set of physical properties $\boldsymbol{X}$.

\begin{table*}[t]
\begin{ruledtabular}
\begin{tabular}{c c c c c c }
Intensive variables & $X$ & $\langle X\rangle$ & $\ointercept(\dot{I}_X)$ & $\soslope(\dot{I}|_X)$ & $\tau_{\mathcal{X}}^{-1} := \Delta\mathcal{Y}$ \\[2pt]
\colrule
- & $\epsilon$ & $U$ & $-\soslope U$ & $\dot{Q}/\Delta\epsilon^2$ & $|\dot Q|/\Delta \epsilon$\\
- & $I = -\ln p$ & $S$ & $-\soslope S$ & $\dot{S}/\Delta I^2$ & $|\dot S|/\Delta I$\\
$\beta$ & $X_\omega:=I-\beta\epsilon$ & $\Phi:= S-\beta U$ & $-\soslope\Phi=\soslope\beta F$ & $\sigma/\Delta X_\omega^2$ & $\sigma/\Delta X_\omega$\\
$\beta$, $\beta P$ & $X_\omega -\beta Pv$ & $\Xi:= \Phi-\beta P \langle v\rangle$ & $-\soslope \Xi=\soslope\beta G$ & $\sigma/\Delta(X_\omega-\beta Pv)^2$ & $\sigma/\Delta (X_\omega-\beta Pv)$\\[2pt]
\colrule
- & $n$ & $\langle n\rangle$ & $-\soslope\langle n\rangle$ & $\dot{N}/\Delta n^2$ & $|\dot N|/\Delta n$\\
$\beta\mu$ & $X_n:=I+\beta\mu n$ & $\Gamma:= S+\beta \mu\langle n\rangle$ & $-\soslope\Gamma$ & $\sigma/\Delta X_n^2$ & $\sigma/\Delta X_n$\\
$\beta$, $\beta\mu$ & $X_\omega+\beta\mu n$ & $\Psi:=\Phi+\beta\mu\langle n\rangle$ & $-\soslope\Psi$ & $\sigma/\Delta(X_\omega+\beta \mu n)^2$ & $\sigma/\Delta (X_\omega +\beta\mu n)$\\
$\{\xi^i\}$ & $X_f: = I- \xi^{i} f_{i}$ & $S - \xi^i\langle f_i\rangle$ & $-\soslope(S-\xi^i\langle f_i\rangle)$ & $(\dot{S}- \xi^i\dot{\mathcal{Q}}_i)/\Delta X_f^2$ &$|\dot{S}-\xi^i\dot{\mathcal{Q}}_i|/\Delta X_f$
\end{tabular}
\end{ruledtabular}
\caption{\label{tab:table1}The linear regression hypothesis, $Y=a+bX$, for the information rate $Y\to -\dot{I}$ with a single predictor variable, $X$, gives an optimal intercept, $\ointercept$, related to Massieu functions. The optimal slope, $\soslope$ determines the fluctuations, $\Delta\mathcal{Y}^2$, in the optimal linear predictor, $\mathcal{Y}=\ointercept+\soslope X$; these fluctuations are the squared speed $\tau_{\mathcal{X}}^{-2}$ at which observables change by one standard deviation $\Delta X$.
Results here are for a pure substance evolving in time and interacting with a reservoir or reservoirs with intensive parameters that are potentially time dependent ($\xi^i=\beta$, $\beta\mu$, $\beta P$, etc.). We only list intensive quantities needed for the regression hypothesis.
Here, we define $k_B=1$, the entropy production rate (for local detailed balance dynamics) $\sigma := \dot{S}-\xi^i\dot{\mathcal{Q}}_i$, and $\dot{\mathcal{Q}}_i$ represents the possible fluxes of energy from non-mechanical forces: heat $\dot{Q}$, chemical work $\dot{W}_{\chem}$, electrical work, magnetic work. Capital Greek letters denote the nonequilibrium analogs of Massieu functions. The Helmholtz free entropy is $\Phi$ and the Planck potential/Gibbs free entropy is $\Xi$. Other Massieu functions, $S - \xi^i\langle f_i\rangle$, can be invented as needed~\cite{callen85}.}
\end{table*}

As in thermodynamics, these \textit{predictor variables}, $\boldsymbol{X}$, can be chosen by the particular experimental setup.
Natural choices are (combinations of) energies $\epsilon$, number of particles, $n$, and the information content or surprisal, $I:=-\ln p$.
The averages of $Y$ and the predictor variables $\boldsymbol{X}$ over the marginal distribution $\boldsymbol{p}$ are nonequilibrium counterparts to equilibrium thermodynamic quantities.
For example, the mean of the discrete energy states $\epsilon$ is $U$ and the mean of the information content, $I:=-\ln p$, is the Shannon entropy, $S$.
Any of these variables we consider, including the intensive parameters of the external reservoirs (if any), may be time dependent.

\addtolength\leftmargini{-0.2in}
\begin{quote}

\noindent\textit{Example 1.0.--} To illustrate the general theory that follows, we include a series of examples focusing on a closed system. The system we take to be out of equilibrium and exchanging energy as heat with an external reservoir at a (possibly time-dependent) inverse temperature $\beta=1/k_BT$.
The displacement from equilibrium could be caused by initial conditions (the removal of a constraint), work done on or by the system by varying the energy states $\epsilon=\epsilon(t)$, or the energy exchanged as heat through thermal contact with a reservoir.
We will analyze the regression of $-\dot{I}=d\ln p/dt$ on the deviations of the distribution from thermal equilibrium $X_\omega= I-\beta\epsilon$.~\footnote{We drop the explicit dependence of variables, such as $d\ln p(z,t)/dt$, on the configuration $z$.}
These quantities can be averaged over the distribution $\{p(z,t)\}$ to define nonequilibrium analogues of the usual thermodynamic properties at equilibrium.
Averaged over the nonequilibrium distribution, $-\langle \dot{I}\rangle =0$ and $\langle X_\omega\rangle=-\beta F$.

\end{quote}

Taking a single predictor variable, $X$, as an example, a natural choice is to add terms, keeping in mind well-known relationships at thermodynamic equilibrium.
At equilibrium, each state $z$ will be occupied according to the Gibbs distribution,
\begin{equation}
  p^{\textrm{eq}}(z|\xi) = Z^{-1} e^{-\xi^{i} f_{i}(z)},
\end{equation}
and have information content $I^\text{eq} = \xi^{i}(t) f_{i}(z) + \ln Z$. The $f_i(z)$ are extensive variables of the system: energy $\epsilon$, number of particles $n$, or volume $V$. The variables $\xi^{i}(t)$ are the corresponding thermodynamic forces: $\beta=1/k_BT$, $\beta\mu$, and $\beta P$.
The ``information deficiency'', $X$, measuring the deviation of the distribution from the equilibrium distribution, is one class of predictor variables:
\begin{equation}
  X_f := X(z|\xi) = I(z)- \xi^{i}(t) f_{i}(z).
\end{equation}
Up to the additive constant $-\xi^i(t)\psi = \ln Z$, they are $I(z)-I(z|\xi)$. Their averages correspond to Massieu functions, $\psi$. Table~\ref{tab:table1} shows representative results for single predictor variables of this form in the modeling of $-\dot{I}$.
As we will see, these predictor variables are natural in that they lead to uncertainty relations and speed limits on the physical quantities, including the heat flux and entropy production rate.

\subsection{Statistical modeling}

With statistical problems of this type, several questions arise: What is the optimal point forecast of $Y$ given a measurement of $\boldsymbol{X}$~\footnote{Alternatively, we might predict the value of a
thermodynamic quantity at a later time $t$ given its measurement at an earlier
time $t_0$.}?
What is their functional relationship?
These questions are common in statistical investigations seeking to establish relationships that make it possible to predict one or more variables from others.
That is, these questions touch more broadly on statistical inference~\cite{kiefer1987introduction} and methods of regression used in supervised machine learning~\cite{friedman2001elements}.

One way forward in answering these questions is to make a regression hypothesis, Fig.~\ref{fig:fig1}(c), and predict a conditional (nonequilibrium) average for $Y$, $\langle Y|\boldsymbol{X}=\boldsymbol{x}\rangle$, and infer its relationship to $\boldsymbol{X}$.
Suppose the true, potentially nonlinear, relationship between $Y$ and the random variables $\boldsymbol{X}$ is $Y|_{\boldsymbol{X}}=f(\boldsymbol{X})$.
A linear regression model hypothesizes the relationship,
\begin{equation}
  Y|_{\boldsymbol{X}} = a + \boldsymbol{b}^\top\boldsymbol{X} + \boldsymbol{\mathcal{E}},
  \label{eq:Mreg}
\end{equation}
and seeks to optimize the unknown coefficients $a$, $\boldsymbol{b}=(b_1,b_2,\ldots,b_N)^\top$ despite the error $\boldsymbol{\mathcal{E}}$ associated with the chosen model.
Linear models simplify the mathematics and, here, they give connections between physical observables. Other regression methods are straightforward generalizations~\cite{kiefer1987introduction}.
By assuming we have the nonequilibrium distribution for the linear model, we are constructing optimal linear predictions.
Optimal predictions neglect any sources of statistical error but also avoid additional modeling assumptions about the measurements, sampling, and noise.

\begin{quotation}

\noindent\textit{Example 1.1.--} Continuing Example 1.0, the linear regression hypothesis is that the rate of change in the information content $\delta\ln p(t)$ over an infinitesimal time interval $\delta t$ is linearly related to the deviations from the canonical equilibrium distribution: $-\dot{I}=a + b (I-\beta\epsilon)$. The model in these examples predicts that the more the nonequilibrium population of state $i$ deviates from the equilibrium population, the higher the rate at which the distribution changes.

\end{quotation}

\subsection{Least squares optimization}

Having chosen a particular statistical model, the next step is to
gauge the quality of the prediction and subsequently identify the optimal prediction of $Y$-- the conditional expectation value $\langle Y|\boldsymbol{X}=\boldsymbol{x}\rangle$.
In regression models, the error
$\boldsymbol{\mathcal{E}}$ measures the component of the response variable $Y$
that is nonlinearly related to $\boldsymbol{X}$.
The variables need not be linearly related, so the error can be large if the relationship is strongly nonlinear.
Taking a variational approach, we choose the mean-squared error as an objective function to find optimal predictions of $Y$ given $\boldsymbol{X}$.
Minimizing the mean-squared error, Fig.~\ref{fig:fig1}(d),
\begin{equation}
  %\min\limits_{a,b}\; \langle\boldsymbol{\mathcal{E}}^\top\boldsymbol{\mathcal{E}}\rangle = \min\limits_{a,b}\; \left\langle [Y -(a + \boldsymbol{b}\cdot\boldsymbol{X})]^\top[Y -(a + \boldsymbol{b}\cdot\boldsymbol{X})]\right\rangle,
  \min\limits_{a,b}\; \langle\boldsymbol{\mathcal{E}}^\top\boldsymbol{\mathcal{E}}\rangle = \min\limits_{a,b}\; \left\langle |Y -(a + \boldsymbol{b}\cdot\boldsymbol{X})|^2\right\rangle,
\end{equation}
gives the regression coefficients that establish the optimal linear relationship between the means of $Y$ and $\boldsymbol{X}$ (Supplementary Material A, SM A): $(\ointercept,\oslope) := \argmin_{a,b}\langle\boldsymbol{\mathcal{E}}^\top\boldsymbol{\mathcal{E}}\rangle$.

First, the optimal ``intercept'' for the regression of $Y$ on $\boldsymbol{X}$,
\begin{equation}
  \ointercept := \ointercept (Y|_{\boldsymbol{X}}) = \langle Y\rangle - \oslope\cdot\langle \boldsymbol{X} \rangle,
\end{equation}
relates the mean $\langle\boldsymbol{X}\rangle$ to $\langle Y\rangle$.
Its physical dimensions are the same as those of $Y$.
Second, the optimal ``slope'',
\begin{eqnarray}
  \oslope := \oslope(Y|_{\boldsymbol{X}}) = \operatorname{cov}[\boldsymbol{X},\boldsymbol{X}]^{-1}\operatorname{cov}[Y, \boldsymbol{X}]
  =: \boldsymbol{\Sigma}^{-1}\boldsymbol{c},
  %\boldsymbol{\beta} = \frac{\operatorname{cov}(\boldsymbol{X},Y)}{\Delta \boldsymbol{X}^2}.
\end{eqnarray} 
is the gradient of the function $(\intercept,\slope)\cdot(1,\boldsymbol{X})$ pointing in the steepest uphill direction in the $N$-dimensional space of predictor variables.
For a single predictor variable, the slope has physical dimensions $Y/X$.
The covariance matrix $\boldsymbol{\Sigma}:=\operatorname{cov}[\boldsymbol{X},\boldsymbol{X}]$ has elements $[\boldsymbol{\Sigma}]_{ij}=\operatorname{cov}(X_i,X_j)$ measuring the correlation between pairs of predictor observables $X_i$ and $X_j$; it must be invertible (i.e., nonsingular~\cite{horn2012matrix}) for the existence of a unique $\oslope$. Physically, we can interpret the magnitude of the slope as a measure of the instantaneous deviation of the system from stationarity; it vanishes at stationary states where $dp(z)/dt = 0$ \,$\forall z$ and $\boldsymbol{c}:= \operatorname{cov}[Y, \boldsymbol{X}]=0$.

For the regression of the information rate $-\dot{I}$, the intercept is directly related to a nonequilibrium average observable and the slope to the corresponding path function.
That is, the optimal regression parameters are composed of thermodynamic quantities, Table~\ref{tab:table1}.
To see this result, we use the fact that the covariance between the surprisal rate $-\dot{I}$ and a variable $X$ under the marginal distribution over configurations is a path function $\operatorname{cov}(-\dot{I},X)=\dot{\mathcal{X}} = \sum_z^M x(z,t) \dot p(z,t)$.
Path functions of this form are measures of linear correlation~\cite{nicholson2020time}.
The optimal slope, $\oslope = \boldsymbol{\Sigma}^{-1}\dot{\boldsymbol{\mathcal{X}}}$, is then directly related to physical quantities.
With a single predictor variable, the magnitude of the slope is inversely related to the time it takes for the path function to evolve the observable $X$ by one standard deviation~\cite{nicholson2020time}: $|\hat b(\dot{I}|_{X})| = \tau_{\mathcal{X}}^{-1}\Delta X^{-1}$~\footnote{We use $\Delta Z^2$ to indicate the variance over the marginal probability distribution over configurations, $\{p_i(t)\}$.}.

\begin{quotation}
\noindent\textit{Example 1.2.--} Minimizing the mean square error gives an optimal intercept that is, in part, a measure of free energy: $\ointercept (\dot{I}|_{I-\beta\epsilon}) = -\soslope\langle I-\beta\epsilon\rangle = \soslope\beta F = -\soslope\Phi$. It is directly related to the nonequilibrium analog of the Helmholtz free entropy $-\Phi=\beta F$ or the Helmholtz free energy, $F=U-TS$~\cite{callen85}.
\end{quotation}
\begin{quotation}
\noindent\textit{Example 1.3.--} The magnitude of the optimal slope is a measure of irreversibility. The expression for the slope contains the covariance between $\dot{I}$ and $X_\omega:= I-\beta\epsilon$.
For the dynamics of a closed system that satisfy local detailed balance, the numerator -- the difference between the Shannon entropy rate $\dot{S}/k_B$ and the heat flux $\beta\dot{Q}$ -- is the entropy production rate, $\sigma=\dot S/k_B - \beta\dot Q \geq 0$ (SM C).
Using $\sigma = \operatorname{cov}(-\dot{I},I-\beta\epsilon)$, the mean-square error is a minimum when $\dot{S}_i$ and fluctuations in the deviations from equilibrium determine the slope,
\begin{equation*}
  \sigma\soslope(\dot{I}|_{I-\beta\epsilon}) %= \frac{}{\operatorname{var}(I-\beta\epsilon)}
  %= \frac{-\operatorname{cov}(\dot{I},I)+\beta\operatorname{cov}(\dot{I},\epsilon)}{\operatorname{var}(I-\beta\epsilon)}\\\notag
  %= \frac{\dot{S}/k_B-\beta\dot{Q}}{\operatorname{var}(I-\beta\epsilon)}
  %= \frac{|\dot S/k_B - \beta\dot Q|}{\Delta(I-\beta\epsilon)^2}
  = \frac{\sigma^2}{\Delta X_\omega^2} = \tau_{\omega}^{-2} \geq 0.
\end{equation*}
Here, we defined the speed $\tau^{-2}_{\omega}$ at which the entropy production rate, $\sigma$, changes the entropy production $\omega=S/k_B - \beta Q$ by one standard deviation, $\Delta X_\omega$.
The magnitude of the slope is increased by the magnitude of the entropy production rate or the speed $\tau_\omega^{-1}$ and suppressed fluctuations in $X_\omega$.
Under these conditions, the second law of thermodynamics is a geometric feature of the linear regression of $\dot{I}$ on $I-\beta\epsilon$. The irreversible increase in entropy associated with spontaneous processes corresponds to a positive slope in this regression hypothesis.
When $d\boldsymbol{p}/dt=0$, the slope $\soslope\to 0$ corresponds to a vanishing speed $\tau^{-1}_{\omega}\to 0$. That is, the slope is zero at equilibrium.

\end{quotation}

From these examples with a single predictor variable, its clear the regression parameters have physical significance.
Given the optimization of a linear model with the mean-squared error led to physical quantities, we can then construct the optimal linear predictor for the relationship between $Y$ and $\boldsymbol{X}$.

\subsection{Optimal linear predictor}

With the coefficients $\ointercept$ and $\oslope$, the optimal linear predictor for the relationship between $Y$ and $\boldsymbol{X}$ is $\mathcal{Y} = \ointercept + \oslope\cdot\boldsymbol{X}$.
Calligraphic typeface $\mathcal{Y}$ will indicate the optimal linear predictor of $Y$ through the regression hypothesis; e.g., $\dot{\mathcal{I}}$ is the optimal linear predictor of $\dot{I}=-d\ln p/dt$.
Equivalently, this optimal predictor of the relationship,
\begin{equation}
  \mathcal{Y}|_{\boldsymbol{X}} = \langle Y\rangle + \delta \boldsymbol{X}^\top\boldsymbol{\Sigma}^{-1}\boldsymbol{c},
  \label{eq:IdealY}
\end{equation}
depends on the correlations between predictor variables through the covariance matrix, $(\boldsymbol{\Sigma})_{ij}=\operatorname{cov}(X_i,X_j)$, and the covariance between each predictor $X_i$ and $Y$ through $\boldsymbol{c}:=\operatorname{cov}(\boldsymbol{X},Y)^T$.
When analyzing the regression of the information rate $-\dot{I}$, the elements of $\boldsymbol{c}$ are physical observables such as the energy exchanged as heat, $\dot{Q}$, and the chemical work, $\dot{W}_\chem$~\cite{nicholson2020time}.
The regression of the information rate also gives an optimal model of the equation of motion for the probability distribution over the statistical manifold, Fig.~\ref{fig:fig1}(b), in terms of the predictor variables $\boldsymbol{X}$.

\begin{quotation}

\noindent\textit{Example 1.4.--} The optimal linear predictor $\mathcal{Y}\to -\dot{\mathcal{I}}|_{I-\beta\epsilon}$ of the relationships between $-\dot{I}$ and $X_\omega$,
\begin{equation}
  -\dot{\mathcal{I}}|_{I-\beta\epsilon} = \left.\frac{d\ln p}{dt}\right|_{\ointercept,\oslope} = \frac{\sigma}{\Delta X_\omega^2}[I-\beta(\epsilon-F)],
\end{equation}
is determined by the entropy production rate $\sigma$ relative to the fluctuations in $X_\omega=I-\beta\epsilon$. For our choice of variables here, $\dot{\mathcal{I}}$ has a mean of zero.

\end{quotation}

The optimal linear predictor has a number of useful properties.
It has a mean $\langle Y\rangle=\langle\mathcal{Y}\rangle$ that is exactly that of the regression variable $Y$ (SM B).
Its variance is related to the optimal slope: $\Delta\mathcal{Y}^2 = \boldsymbol{c}^\top\oslope$.
Rearranging Eq.~(\ref{eq:IdealY}), the prediction uncertainty, $\delta\mathcal{Y}=\delta \boldsymbol{X}^\top\boldsymbol{\Sigma}^{-1}\boldsymbol{c}$,
%\begin{equation}
%  \delta\mathcal{Y} = \sum_{i,j}^N\delta X_i\operatorname{cov}(X_i,X_j)\operatorname{cov}(X_j,Y),
%\end{equation}
is zero when $Y$ is a linear function of $\boldsymbol{X}$.
Otherwise, the uncertainty is nonzero and an optimal prediction for the chosen regression hypothesis and objective function.
In other words, the linear regression hypothesis measures the amount of linear correlation between $Y$ and $\boldsymbol{X}$ through the slope, even when $Y$ and $\boldsymbol{X}$ are nonlinearly related.

\section{Thermodynamic uncertainty relations and speed limits}
\subsection{Thermodynamic uncertainty relations}

While the linear model is chosen to be optimal, we still need to assess its quality since the relationship between $Y$ and $\boldsymbol{X}$ may not actually be linear.
The variance of the optimal linear predictor is one way to make this assessment, and, as we will see, leads to thermodynamic uncertainty relations and, for $Y\to -\dot{I}$, speed limits on observables.
That is, there are the thermodynamic consequences of predicting the optimal relationships and measuring the amount of linear correlation between physical and information-theoretic observables in nonequilibrium processes.

Using the optimal regression coefficients $\ointercept$ and $\oslope$, the \textit{minimum} prediction errors associated with the regression hypothesis,
$\langle\boldsymbol{\mathcal{E}}^\top\boldsymbol{\mathcal{E}}\rangle|_{\ointercept,\oslope} = \Delta Y^2[1-\rho_{Y,\boldsymbol{X}}^2]\geq 0$, are determined by the correlations between $\boldsymbol{X}$ and $Y$.
Here, $\rho_{Y,\boldsymbol{X}}^2 =1- \Delta\mathcal{Y}^2/\Delta Y^2$ is a form of the Pearson correlation coefficient. Rearranging the minimum mean-squared error gives (SM B) a multivariate uncertainty relation for any number of thermodynamic observables:
\begin{equation}
  \rho^2_{Y,\boldsymbol{X}} = \boldsymbol{c}^\top \Delta Y^{-2}\boldsymbol{\Sigma}^{-1}\boldsymbol{c} %= \boldsymbol{c}^\top[\operatorname{var}(Y)]^{-1}\boldsymbol{\beta} 
\leq 1.
  \label{eq:gur}
\end{equation}
The Pearson correlation coefficient, $\rho_{Y,\boldsymbol{X}}$, is a measure of the linear correlation between $\boldsymbol{X}$ and $Y$.
For example, if there is a single predictor variable, $\operatorname{cov}^2(X,Y)\leq \Delta X^2\Delta Y^2$.
This relation for ``prediction errors'' associated with the linear hypothesis for the thermodynamic regression of $Y$ on $\boldsymbol{X}$, Eq.~(\ref{eq:Mreg}), follows from the choice of the linear hypothesis and the mean-squared error objective function.
The prediction errors are uncorrelated with $\boldsymbol{X}$, $\operatorname{cov}(\boldsymbol{X},Y-\mathcal{Y})=0$.

\begin{quotation}
\noindent\textit{Example 1.5.--} Defining $X_\omega$, the Pearson correlation coefficient,
\begin{eqnarray*}
  \rho^2_{-\dot{I},X_\omega} = %\boldsymbol{c}^\top\boldsymbol{\beta}(\dot{I}|_{X_\epsilon})\Delta\dot{I}^{-2} = 
  \frac{\sigma^2}{\Delta\dot{I}^2\Delta X_\omega^2} &\leq& 1,
\end{eqnarray*}
measures the quality of the linear hypothesis and leads to the thermodynamic uncertainty relation for the entropy production rate: $\sigma \leq \Delta\dot{I}\Delta X_\omega$.

\end{quotation}

The model for observables, such as those listed in Table~\ref{tab:table1}, is a choice one must make in analyzing a particular nonequilibrium process.
Regardless of the choice, the associated multivariable uncertainty relation is not specific to a particular nonequilibrium current or the entropy production rate.
Nor does it require any assumptions about the stochastic
dynamics, the proximity to equilibrium, the size of the system, or the protocol
driving the system out of equilibrium.
Since it is based on regression theory, it is important to point out that it does not require an assumption about the form of the marginal distribution, $\{p_i\}$, or the fluctuations about the optimal predictor. 

\subsubsection{Fluctuations of the optimal linear predictor and saturation of the uncertainty relation}

The optimal linear predictor is defined so that its fluctuations saturate the uncertainty relation.
That is, the variance of the optimal linear predictor $\mathcal{Y}$ gives %$\operatorname{var}(\mathcal{Y}) \operatorname{var}(\boldsymbol{X}) =
%\operatorname{cov}(\boldsymbol{X},Y)^2$ or
\begin{eqnarray}
  \boldsymbol{c}^\top\,\Delta \mathcal{Y}^{-2} \boldsymbol{\Sigma}^{-1} \boldsymbol{c} = 1,
  \label{eq:YVar}
\end{eqnarray}
which is an uncertainty equality for the fluctuations of the optimal linear predictor of $Y$.
The correlation coefficient is also related to the variance of the optimal predictor, $\Delta \mathcal{Y}^2/\Delta Y^2 =
\rho^2_{Y,\boldsymbol{X}}$, implying $-\Delta Y \leq \Delta\mathcal{Y} \leq \Delta Y$ (SM B).
%The variance of the optimal predictor saturating these uncertainty relations prompts a closer inspection of its properties.
%For example, the mean of $\mathcal{Y}$ is such that $\langle\mathcal{Y}\rangle=\langle Y\rangle$.
Taking the deviations of a single predictor, $\delta X = X
- \langle X \rangle$, makes it clearer that the optimal predictor can be rearranged into a form,
\begin{equation}
  \label{eq:onsagerlike}
  \frac{\delta\mathcal{Y}}{\delta X} = \frac{\Delta\mathcal{Y}^2}{\operatorname{cov}(X,Y)},
\end{equation}
reminiscent of Onsager's regression hypothesis~\cite{onsager1931reciprocal}. An important distinction, though, is the lack of any assumptions here about linear response or the magnitude of the perturbations from equilibrium.
Our relation instead derives from the optimal linear prediction of a potentially nonlinear relationship between $Y$ and $\boldsymbol{X}$ (here $X$) at any moment in time during a nonequilibrium process.

The rates of change in thermodynamic observables caused by path functions $\dot{\mathcal{X}}$ can constrain the relationship between equation of motion on the statistical manifold and measured observables.
Choosing $Y=-\dot{I}$, we can identify the path function $\dot{\mathcal{X}}:=\operatorname{cov}(X,-\dot{I})$, substitute in the expression above, and solve:
\begin{equation}
  \dot{\mathcal{X}} \frac{\delta \mathcal{Y}}{\delta X} = \rho_{-\dot{I},X}^2 I_F \leq I_F.
\end{equation}
Because the right hand side is non-negative, the three variables on the left can have two sets of signs: either all three are positive or one is positive and two are negative.

For example, if the predictor variable is $X\to\epsilon$, then $\operatorname{cov}(\epsilon,-\dot I) = \dot Q$.
When $\dot Q> 0$, both deviations must be positive, $\operatorname{sgn}(\delta \mathcal{Y}) = \operatorname{sgn}(\delta X)$.
When $\dot Q < 0$, then $\operatorname{sgn}(\delta \mathcal{Y}) = -\operatorname{sgn}(\delta X)$.
So, knowing the change in heat and predictor variable fixes the sign of the deviations in the optimal predictor of $d_t\ln p$.

If $\dot{\mathcal{X}}\geq 0$, then it follows that $\operatorname{sgn}(-\dot{I})=\operatorname{sgn}\delta X$. This relationship means that positive (negative) deviations $\delta X$ correspond to positive (negative) intrinsic rates $-\dot{I}=d_t\ln p$. The nonnegativity of $\dot{\mathcal{X}}$ then constrains the relationship between the intrinsic rate at which probability distributions evolve $-\dot{I}$ and the deviations of observables from their nonequilibrium averages $\delta X$.

\begin{quotation}
\noindent\textit{Example 1.6.--} 
Again take the single predictor variable, $X_\omega = I - \beta\epsilon$. The covariance of this variable with $-\dot{I}$ is the entropy production, $\sigma = \operatorname{cov}(-\dot I,I-\beta\epsilon)$. When the time-information uncertainty relation saturates, $\rho^2 = 1$, the probability over configurations evolves as $-\dot{\mathcal{I}} = \sigma \delta X_\omega/\Delta X_\omega^2$ and the entropy production rate is exactly $\sigma = \Delta\dot{I}\Delta X_\omega$.
For a system with a dynamics satisfying local detailed balance, $\sigma\geq 0$~\cite{esposito2011second}. Our result above then means that $-\delta \mathcal{I}$ and $\delta X_\omega$ share a sign.
Using $\rho^2\leq 1$ and $I_F\geq 0$ in Eq.~(\ref{eq:onsagerlike}) gives:
\begin{equation*}
  0\leq \frac{\sigma}{I_F} = \rho^2_{-\dot{I},X_\omega}\frac{\delta X_\omega}{\delta \mathcal{Y}} \leq \frac{\delta X_\omega}{\delta \mathcal{Y}},
\end{equation*}
when the system is away from a stationary state, $I_F>0$, and $\dot{\mathcal{I}}\neq 0$.
%If $\dot{S}_i\geq 0$ then $\delta X/\delta \mathcal{Y}\geq 0$.
%\begin{equation}
%  \dot{S}_i = \Delta \mathcal{Y}^2\left(\frac{\delta X}{\delta\mathcal{Y}}\right) = \rho^2_{\boldsymbol{X}Y}I_F\left(\frac{\delta X}{\delta\mathcal{Y}}\right)
%\end{equation}
Since $\operatorname{sgn}(-\dot{\mathcal{I}})=\operatorname{sgn}\delta X$, positive (negative) deviations $\delta X(z)$ correspond to positive (negative) intrinsic rates $d_t\ln p(z)$\,$\forall z$ across the set of configurations.
For system interacting with a non-ideal reservoir, such as one that is of comparable size to the system $\sigma$ can become negative, for example, due to correlations between the reservoir and system~\cite{esposito2010entropy}. A similar idea emerges here in a different context, and, if $\sigma \leq 0$, then this correlation means that $\delta X_\omega$ and $-\mathcal{I}$ must have opposite signs.

\end{quotation}

\subsection{Thermodynamic speed limits and time-information uncertainty relations}

In thermodynamics, experimental conditions dictate the natural variables $\boldsymbol{X}$ and appropriate thermodynamic potentials $Y$~\cite{callen85}.
 Away from equilibrium, the choice is less clear.
How the uncertainty bounds on one observable relate to other observables has been explored using parameter estimation (in the case of fixed parameters) through the multivariate Cramer-Rao
bound~\cite{dechant2018multidimensional}.
How uncertainty bounds couple and compete for driven systems is still an open question. We recently discovered a family of single-variable time-information uncertainty relations that can be cast as thermodynamic speed limits~\cite{nicholson2020time}.
Using the theory above, we can show these results are part of a formalism for the regression of information.
Recognizing this broader framework, we can derive multidimensional generalizations that apply to multiple observables and coupled flows of energy, entropy, and matter.
Looking back, the example we have used to illustrate the theory above is one of a number of possible choices for the predictor variables of a linear model for $Y\to -\dot{I}$.
Before discussing the single and multivariable version of the time-information uncertainty relation, we first define the fluctuations in information content that accompany nonequilibrium processes.

\subsubsection{Information fluctuations and intrinsic speed}

Observables evolve at a speed $\tau_{\mathcal{X}}^{-1}$ that depends on the sensitivity of the observable $X$ to changes in the distribution over configurations.
This speed, however, is bounded by the speed at which probability distributions evolve by a distance $ds = \sqrt{I_F} \,dt$~\cite{heseltine2016novel},
The square root of the Fisher information~\cite{FisherInfo}, $\sqrt{I_F}$, has been used as an intrinsic speed in both quantum~\cite{StatDistQuantum,DistQuantumTwo,Taddei13} and classical~\cite{SalamonB83,SalamonNB85,MLthermo,flynn2014measuring,nichols2015order} settings.
The statistical distance $ds$,
\begin{align}
  ds^2 &= \sum_{k,l} \frac{d\lambda^k}{dt}g_{kl}\frac{d\lambda^l}{dt}dt^2 = I_F \,dt^2,
  \label{eq:distance}
\end{align}
which can also be expressed in terms of the Fisher metric,
\begin{align}
  g_{kl} = \left\langle\frac{\partial \ln p(z)}{\partial\lambda_k}\frac{\partial \ln p(z)}{\partial\lambda_l}\right\rangle,
  \label{eq:IFMet}
\end{align}
measures the distinguishability between $\boldsymbol{p}(t)$ and
$\boldsymbol{p}(t+dt)$~\cite{heseltine2016novel,StatDistQuantum}. From the physical
dimensions, $\sqrt{I_F}$ is a speed relating this dimensionless measure of
distance $ds$ to an infinitesimal increment of time $dt$. The Fisher information~\cite{FisherInfo} is also a measure of fluctuations in the surprisal rate:
\begin{equation}
\label{eq:Fisher}
  I_F := \sum_{z} p(z)\left(\frac{d\ln p(z)}{dt}\right)^2 = \Delta {\dot{I}}^2 =: \frac{1}{\tau^2} \geq 0.
\end{equation}
For systems that are not in stationary states, fluctuations in the information content, $1/\Delta\dot{I} =
1/\sqrt{I_F} =: \tau$, set an intrinsic timescale for the evolution of the
probability distribution~\cite{flynn2014measuring,nichols2015order,kim2016geometric}.

\subsubsection{Time-information uncertainty relations}

The regression hypothesis for information rates leads us to a speed limit on predictor variables.
With $Y=-\dot{I}$, any choice of predictor variables gives the multivariable \textit{time-information uncertainty relation},
\begin{eqnarray}
  \rho^2 = \boldsymbol{c}^\top I_F^{-1}\oslope &\leq& 1\\\notag
  \boldsymbol{c}^\top \oslope &\leq& I_F = \tau^{-2},
\end{eqnarray}
in which the fluctuations in $Y$ are the Fisher information $I_F = \Delta \dot{I}^{2}$ parameterized by time.
The general framework we have laid out then includes a class of time-information uncertainty relations that set speed limits on the evolution of arbitrary observables in thermodynamics~\cite{nicholson2020time}.
It is the fluctuations in the optimal predictor that set an intrinsic speed for the evolution of observables.
To illustrate, letting $\boldsymbol{X}\to X$ (a single observable), we find
$\mathcal{Y}=\langle Y\rangle + \delta X\Delta\mathcal{Y}^2$ and see that
$\langle\mathcal{Y}\rangle = \langle Y\rangle$.
We can use that the fluctuations, $\Delta \mathcal{Y}^2=\dot{\mathcal{X}}^2/\Delta X^{2}$, set the intrinsic speed at which the observable $X$ changes by a statistically distinguishable amount~\cite{nicholson2020time} where the rate of change is of the form:
$\mathcal{\dot{X}} = \sum_z x(z,t) \dot p(z,t)$.
The timescale $\speedtime$ for these observables to evolve to a statistically distinguishable value,
\begin{equation}
  \speedtime^2 = \frac{\mathcal{\dot{X}}^2}{\Delta X^2} = \frac{1}{\Delta\mathcal{Y}^2}
\end{equation}
we can now see is inversely related to the fluctuations in the optimal linear
predictor. Here, $\mathcal{X} = \int_{t_0}^{t_f}\mathcal{\dot{X}}\,dt$; for example, $\omega = \int_{t_0}^{t_f}\sigma\,dt$. 

If the response variable is $Y\to\dot{I}$, then $\langle \dot{I}\rangle =
\langle\mathcal{Y}\rangle = \langle Y\rangle = 0$. The optimal linear predictor
is $\mathcal{Y}=\delta X\,\dot{\mathcal{X}}^2/\Delta X^{2}=\delta
X/\speedtime^{2}$. And, in this case, it is only a function of its fluctuations
$\mathcal{Y}=\delta X\Delta\mathcal{Y}^2$, which means it is also only a
function of the intrinsic speed $\mathcal{Y}=\delta X/\speedtime^2$. The
variable $Y\to\dot{I}$ has fluctuations measured by the Fisher information
$\Delta Y^2=\Delta\dot I=I_F=\tau^{-1}$, which sets a speed limit on the
intrinsic speed of thermodynamic observables:
\begin{equation}
  \Delta \mathcal{Y} = \speedtime^{-1} \leq \tau^{-1} = \sqrt{I_F}.
\end{equation}
That is, the Fisher information sets a speed limit on the fluctuations in the optimal linear predictor for the regression of $\boldsymbol{X}$ on $\dot{I}$ (SM D).
To operate at the speed limit, a system must incur fluctuations $\Delta
\mathcal{Y}^2$ in the optimal linear predictor that are precisely $I_F$.

\begin{quotation}
\noindent\textit{Example 1.7.--} We define the speed at which the entropy production rate, $\sigma := \dot{S}/k_B-\beta\dot{Q}$, changes the entropy production $\omega=S/k_B - \beta Q$ by one standard deviation: $\tau^{-2}_{\omega} := \sigma^2/\Delta X_\omega^2$. Now, using the Fisher information $\sqrt{I_F}=\tau^{-1}$, the Pearson correlation coefficient can be expressed as the time-information uncertainty relation,
\begin{eqnarray*}
  \tau_{\omega} \sqrt{I_F} \geq 1,
\end{eqnarray*}
or as the speed limit $\tau_{\omega}^{-1}\leq \tau^{-1}$.

\end{quotation}

When $\dot{\mathcal{X}}=\langle X\rangle$, the time-information uncertainty relation simplifies. To see how, consider a single predictor variable. The mean of observables obey an equation of motion analogous to the Ehrenfest's equation in quantum mechanics: $d_t\langle X\rangle = -\operatorname{cov}(\dot{I},X) + \langle d_t X\rangle$~\cite{nicholson2020time}. For example, if $X$ is energy, the second term is the work done on or by the system. Neglecting this term, the speed limit reduces to the Cram{\'e}r lower bound: $\tau^{-2}_{\mathcal{X}} = [d_t\langle X\rangle]^2/\Delta X^2\leq \tau^{-2}$~\cite{hasegawa2019uncertainty,ItoDechant2020}.
Using the slope $\soslope = [\tau_{\mathcal{X}}\Delta X]^{-1} = d_t\langle X\rangle/\Delta X^2$, the optimal linear predictor, $\dot{\mathcal{I}} = \delta X\soslope = \delta X \dot{\mathcal{X}}/\Delta X^2$, becomes $\dot{\mathcal{I}} = \delta X d_t\langle X\rangle/\Delta X^2$. That is, the equation of motion on the statistical manifold predicted by the regression hypothesis is determined by the deviations of each observable $X$ from its mean $\langle X\rangle$ and the rate of change in the mean predictor variables relative to the fluctuations.

%\sbn{Langevin dynamics and Markovian dynamics can be written as linear regression hypothesis?}
%Statistical Mechanics can be seen as a collection of Mathematical puzzle pieces in which we are trying to construct the complete picture. In equilibrium this picture is mostly finished. Macroscopic observations are explained through the behavior of microscopic configurations. ect. 

%Away from equilibrium this picture is still very much a work in progress. Fundamental results such as the Clausius inequality cite[] and efficiency of processes has given hints of the relations between heat work and the evolution of thermodynamic observables. More recently Stochastic Thermodynamics cite[] has shown how information and thermodynamics are intimately linked. A consistent theme has been that of fluctuations and how their reach extends beyond equilibrium. Onsager's regression hypothesis [], explains the relation between fluctuations around equilibrium and driven evolutions away from it. This mainstay of modern statistical mechanics was proven through the advent of fluctuation dissipation relations, again explaining the observations we make at a macroscale through bounds regarding fluctuations at a microscale. 

%The thermodynamic consequences of $\mathcal{Y}$ are that the optimal linear
%predictor allows one to link the time information uncertainty relation to
%Onsager's regression hypothesis.

\begin{figure*}[t]
\centering
\includegraphics[width=0.9\textwidth]{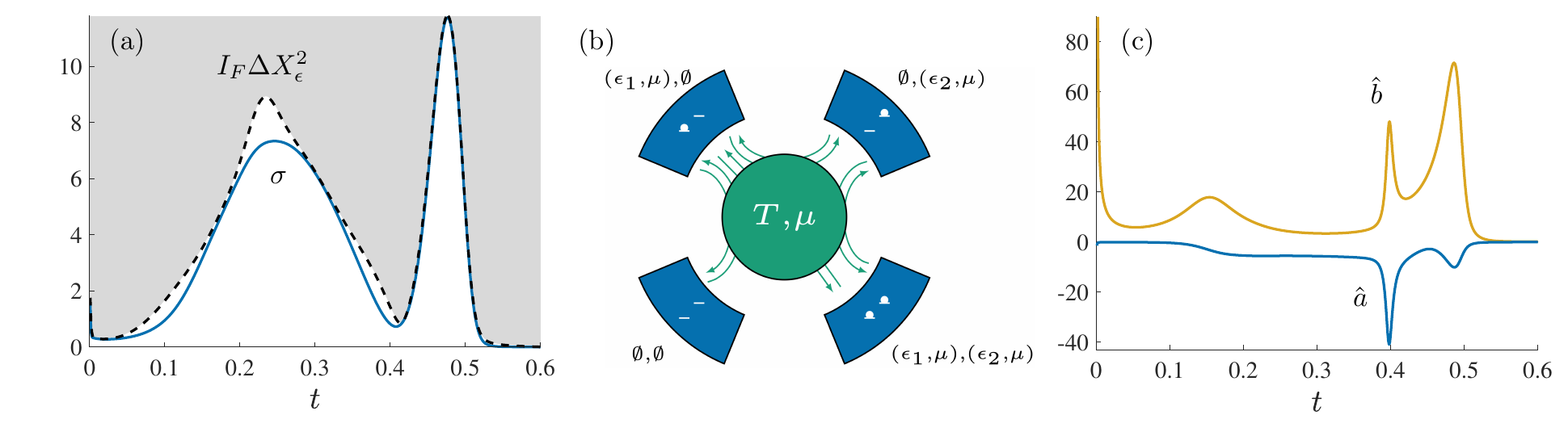}
\caption{\label{fig:fig2} (a) Time-information uncertainty relation for the entropy production rate as a function of time for a driven process: $\sigma=\dot{S}-\beta\dot{Q}\leq I_F\Delta X_\omega^2$. (b) System represented by nodes and transitions by edges. Each outer node is a state of the system. Transitions between states exchange heat and or matter to the reservoir (center) at a fixed temperature $T$ and chemical potential $\mu$. (c) Optimal coefficients $\ointercept$ and $\oslope$ for the linear regression of $-\dot{I}$ as a function of time. The optimal slope $\hat{b}$ is positive at all times as a consequence of the nonnegativity of the entropy production, the second law of thermodynamics.}
\end{figure*}

\subsubsection{Correlations and the number of predictor variables}

The single predictor variable presented in the examples so far can be modified by adding terms to the information deficiency $I-\xi^if_i$.
However, our main results also apply to the linear regression of $Y$ on multiple predictor variables $\boldsymbol{X}$.
The distinction between these cases amounts to knowledge of the additional correlation between the predictor variables $\{X_i\}$.
What then is the effect of this additional correlation on the uncertainty relations and speed limits here?

To answer this question, we consider the bivariate and univariate linear regression of $-\dot{I}$.
Considering the regression of $-\dot{I}$ on $X_1$ and $X_2$, we
find the optimal intercept
\begin{equation}
  \ointercept = -\oslope\cdot\left\langle\begin{matrix} X_1\\ X_2 \end{matrix}\right\rangle
\end{equation}
and the slope $\oslope=\boldsymbol{\Sigma}^{-1}\boldsymbol{c}$,
\begin{equation}
  %\boldsymbol{\beta}=\operatorname{cov}(\boldsymbol{X},\boldsymbol{X})\begin{pmatrix} \operatorname{cov}(-\dot{I},I-\beta\epsilon) \\ \operatorname{cov}(-\dot{I},I-\beta\epsilon) \end{pmatrix}
  %\begin{pmatrix} \Delta X_\epsilon^2 & \operatorname{cov}(X_\epsilon,X_n) \\ \operatorname{cov}(X_n,X_\epsilon) & \Delta X_n^2 \end{pmatrix}
  %\begin{pmatrix} \dot{S}/k_B-\beta\dot{Q} \\ \dot{S}/k_B-\beta\dot{W}_{\chem} \end{pmatrix}.
  \oslope = \begin{pmatrix} \Delta X_1^2 & \operatorname{cov}(X_1,X_2) \\ \operatorname{cov}(X_2,X_1) & \Delta X_2^2 \end{pmatrix}^{-1}
  \begin{pmatrix} \dot{\mathcal{X}}_1 \\ \dot{\mathcal{X}}_2 \end{pmatrix}.
\end{equation}
The inverse of the covariance matrix $\boldsymbol{\Sigma}$ can be calculated explicitly. For any square $n \times n$ matrix, $\mathbf{A}$:
$\mathbf{A}^{-1}=|\mathbf{A}|^{-1} \mathbf{C}^{\top}$~\cite{horn2012matrix}.
We then have the inverse of the covariance matrix:
\begin{equation}
  %\boldsymbol{\Sigma}_{\rho}^{-1}=
\frac{1}{|\boldsymbol{\Sigma}|}\left[
  \begin{array}{cc}
    \Delta X_{2}^2 & -\operatorname{cov}\left(X_{1}, X_{2}\right) \\
    -\operatorname{cov}\left(X_{1}, X_{2}\right) & \Delta X_{1}^2
  \end{array}\right]
\end{equation}
in terms of its determinant,
\begin{equation}
  |\boldsymbol{\Sigma}|=\left|\operatorname{cov}(\boldsymbol{X}, \boldsymbol{X})\right| = \Delta X_{1}^2 \Delta X_{2}^2-\operatorname{cov}(X_{1}, X_{2})^{2},
\end{equation}
measuring the correlations between properties 1 and 2. 

From the multivariate correlation coefficient $\rho^2\leq 1$, the Fisher information bounds the coupled rates (e.g., of energy and matter flow):
\begin{equation}
  %\rho^2I_F\left|\operatorname{cov}\left(\boldsymbol{X}, \boldsymbol{X}\right)\right| = 
  \sqrt{\frac{\dot{\mathcal{X}}_{1}^2 \Delta X _{2}^2+\dot{\mathcal{X}}_{2}^2 \Delta X_{1}^2-2 \operatorname{cov}(X_{1}, X_{2}) \dot{\mathcal{X}}_{1}\dot{\mathcal{X}}_{2}}{\Delta X_{1}^2 \Delta X_{2}^2-\operatorname{cov}(X_{1}, X_{2})^{2}}}
  \leq \tau^{-1}.
\end{equation}
The fluxes $\dot{\mathcal{X}}_{1}$ and $\dot{\mathcal{X}}_{2}$ are the weights in the weighted sum of the variances in equilibrium deviations -- geometrically, a weighted cosine law.
This thermodynamic uncertainty relation bounds the individual fluxes (1 and 2) caused by nonequilibrium currents.
If the two predictor variables $X_1$ and $X_2$ are uncorrelated, $\operatorname{cov}(X_1, X_2)=0$.
%\begin{equation}
%  \dot{S}_{i,\epsilon}^2 \Delta X_{n}^2+\dot{S}_{i,n}^2 \Delta X_{\epsilon}^2
%  \leq I_F\Delta X_{\epsilon}^2 \Delta X_{n}^2
%\end{equation}
%and
%\begin{equation}
%  \sqrt{\frac{\dot{\mathcal{X}}_{1}^2}{\Delta X_{1}^2} + \frac{\dot{\mathcal{X}}_{2}^2}{\Delta X_{2}^2}}
%  \leq \tau^{-1}.
%\end{equation}
In this case, defining $v_j:= \dot{\mathcal{X}}_{j}/\Delta X_j$ (complementing the speed $\tau_j^{-1}:=|\dot{\mathcal{X}}_j|/\Delta X_j$), the sum of their squares obey the speed limit set by $v= \sqrt{I_F} = \tau^{-1}$:
\begin{equation}
  \sqrt{v_1^2 + v_2^2} \leq v.
\end{equation}
We can weaken this speed limit so that $\sqrt{N^{-1}(v_{1}^{2} + v_{2}^{2})} \leq v$ and $v$ upper bounds the r.m.s.\ predictor speeds.
Generalizing for $N$ independent predictor variables $\boldsymbol{X}$,
\begin{equation}
  \boldsymbol{v}_\text{rms} \leq \|\boldsymbol{v}_x\|_2 \leq v.
\end{equation}
The square root of the Fisher information is a speed limit on the sum of square speeds for any number of uncorrelated predictor variables.

\subsubsection{Model system with energy and matter flux}

To compare the thermodynamic speed limits that result from one and two predictor variables, consider a system with a finite number of discrete states that can exchange both energy and particles with external reservoirs.
Particles can occupy two states with energies $\epsilon_1$ and $\epsilon_2$, Fig.~(\ref{fig:fig2}b).
Assuming Fermi statistics, there are four configurations of the system, $s_i = [\{\emptyset,\emptyset\},\{(\epsilon_1, \mu),\emptyset\},\{\emptyset,(\epsilon_2,\mu)\},\{(\epsilon_1, \mu),(\epsilon_2,\mu)\}]$.
Each configuration has an associated energy, $E_i = \sum_{j\in s_k} \epsilon_j$,  and chemical work, $G_i = \sum_{j\in s_k} g_j$.
Adapting the model dynamics from Ref.~\cite{horowitz2019thermodynamic}, we define the continuous-time master equation dynamics:
\begin{equation}
  W_{ij}= \left \{
  \begin{aligned}
  &\alpha e^{-\beta(\delta E_{ij} - \delta G_{ij})} && \text{if}\ j > i \\
  & \alpha && \text{if}\ j < i \\
  & -\textstyle\sum_{i\neq j} W_{ij}, &&\text{if}\ i = j
  \end{aligned} \right.
\end{equation}
where $\delta E_{ij} = E_i - E_j$ and $\delta G_{ij} = G_i - G_j$ with inverse temperature $\beta=4$ and rate parameter $\alpha=4$. 

\begin{figure*}[!t]
\centering
\includegraphics[width=0.9\textwidth]{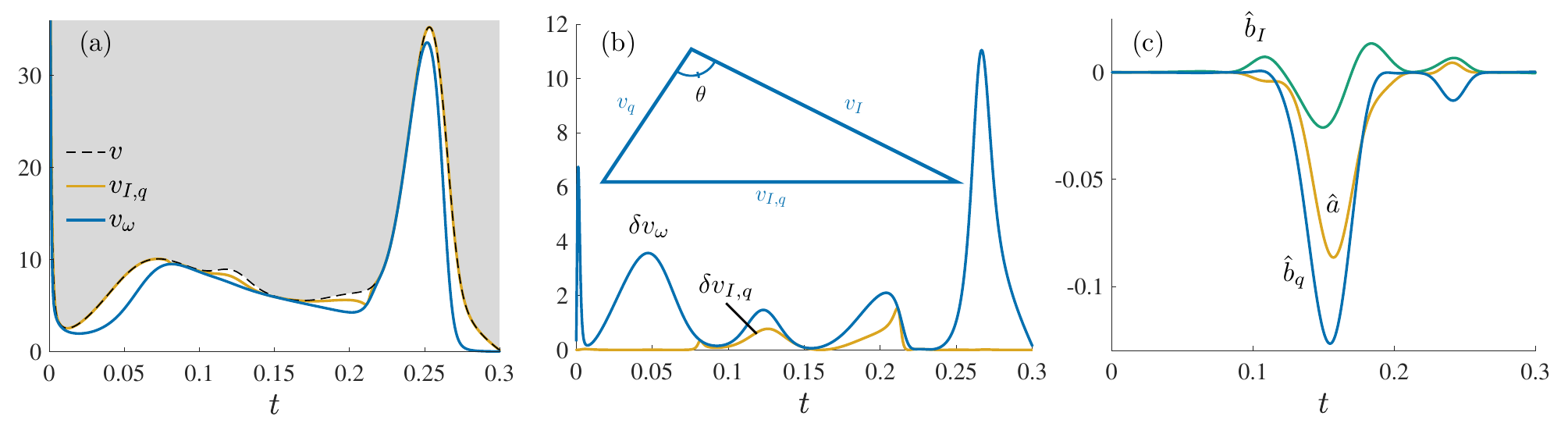}
\caption{\label{fig:fig3} (a) Speed limit on entropy production rate set by the Fisher information (dashed). Univariate speed of entropy production (blue) and tighter bivariate speed (yellow) (b) Difference in speeds, $\delta v$, for univariate bound $\delta v_{\omega} = \sqrt{I_F} - v_{\omega}$ (blue) and bivariate bound $\delta v_{I,q} = \sqrt{I_F} - v_{I,q}$ (yellow). The difference is pronounced when the Fisher information changes rapidly. (c) The three optimal coefficients for the bivariate regression hypothesis of $-\dot{I}$ on $\hat{a}+\soslope_II+\soslope_q\beta q$ change most significantly when the bivariate speed deviates from $\sqrt{I_F}$.}
\end{figure*}

The system can be externally driven by the external environment through the inverse temperature $\beta$, chemical potential $\mu$, or work by varying the energy levels.
Fixing $\mu = 0.9$, we drive the system by the ground state energy according to the protocol $\epsilon_1(t) =1.2 + \cos(2\pi ft)$ with $f=1$ at fixed $\beta=1/k_BT$.
The entropy production rate is the difference between the entropy rate $\dot S$ and the change in heat $\beta\dot Q = \sum_i p_i(t)q_i$, where, $q_i = E_i - G_i$.

For the regression of $-\dot{I}$, we choose the energy $\epsilon_i$ of each state $i$ and the local chemical work $g_i = \mu n_i$ where $\mu$ is the chemical potential and $n_i$ is the state occupation number.
As a point of comparison, Figure~(\ref{fig:fig2}a) shows results for the regression of $-\dot{I}$ on $I-\beta q$; the entropy production rate is bounded by $\tau^{-1}_{\omega}= \sigma/\Delta(I - \beta q) \leq \tau^{-1}$.
Alternatively, instead of measuring the fluctuations in one predictor variable $I - \beta q$, we can treat $I$ and $\beta q$ as separate predictor variables and analyze how this additional information affects the multivariate uncertainty relation.

Figure~(\ref{fig:fig3}c) shows the coefficients,
\begin{equation}
\ointercept = -\,\oslope^\top\langle \mathbf{X}\rangle = -[\dot{S}/k_B,\beta \dot{Q}] \cdot \oslope
\end{equation}
\begin{equation}
\oslope = \begin{bmatrix} \soslope_I \\ \soslope_q\end{bmatrix} = \boldsymbol{\Sigma}^{-1}\begin{bmatrix} \dot S/k_B \\ \beta \dot Q\end{bmatrix},
\end{equation}
of the optimal linear predictor as a function of time.
As we have shown, the multivariate uncertainty relation can be cast as a speed limit:
\begin{equation}
  \frac{\beta^2}{|\boldsymbol{\Sigma}|}\left(\dot S^2\Delta X_q^2 +\dot Q^2\Delta X_i^2 - 2\dot S\dot Q\operatorname{cov}(I,q)\right) \leq I_F = \tau^{-2}
\label{eq:vel1}
\end{equation}
The determinant of the covariance matrix, $|\boldsymbol{\Sigma}| = \beta^2\Delta X_q^2\Delta X_I^2 - \beta^2\operatorname{cov}(I,q)^2$, measures the correlations between $I$ and $q$.

The inner product form of the covariance suggests expressing it as an angle between $\Delta X_I$ and $\Delta X_q$:
\begin{equation}
%\operatorname{cov}(X_I,X_q) = \bra{\delta X_\epsilon\,\sqrt{p}}\ket{\delta X_n\,\sqrt{p}} = \Delta X_I \Delta X_q \cos \theta
\operatorname{cov}(X_I,X_q) =\Delta X_I \Delta X_q \cos \theta.
\end{equation}
Pulling $\Delta X_I\Delta X_q$ out of the determinant and defining $v_Q = \dot Q/\Delta q$ and $v_I = \dot S/\Delta I$, Eq.~(\ref{eq:vel1}) can be seen as a cosine law, Fig.~(\ref{fig:fig3}b):
\begin{equation}
  v_{I,q} := \sqrt{v_I^{-2} + v_q^{-2} - 2v_I^{-1}v_q^{-1}\cos \theta}.
  \label{eq:vel_Iq}
\end{equation}
The multivariate speed limit is then:
\begin{equation}
\tau^{-1} \geq |\sin\theta|^{-1}\tau_{I,q}^{-1}.
\end{equation}
Figure~(\ref{fig:fig3}a) shows that $\tau_{I,q}^{-1}$ is at least as tight as the single predictor speed, $\tau_{\omega} = \Delta(I - \beta q)/\sigma$.
Figure~(\ref{fig:fig3}b) illustrates that when the Fisher information is changing rapidly, the difference between the univariate and bivariate speeds can be substantial, here differing by as much as an order of magnitude.

\medskip

\section{Conclusions}

%Thermodynamic uncertainty relations are tradeoffs between the precision of nonequilibrium currents and dissipation.
The time-information uncertainty relation puts bounds on the speed evolution of thermodynamic observables and path functions, including the entropy production.
Here, we have shown that multidimensional uncertainty relations and speed limits are measures of correlations that derive from optimal linear models for the equation of motion on statistical manifolds.
%The theoretical framework here establishes links between stochastic thermodynamics, information geometry, and regression.
The regression of the rate of information content appears to be unique in that the choice of the predictor variables leads to the time-information uncertainty relation and associated speed limit, with the speed on nonequilibrium observables evolve set by the fluctuations in the optimal linear predictor.
Within this formalism, the second law of thermodynamics is a geometric condition -- a nonnegative slope -- in the linear regression of the information content rate on the information deficiency of the nonequilibrium state. Stationary processes, equilibrium and nonequilibrium steady states, have a vanishing slope.
%Regression of $\dot I$ on the energy $\epsilon$ puts a speed limit on heat flux while regression on $I-\beta\epsilon$ leads to the corresponding speed limit on entropy production.
In sum, these results suggest a broader framework for statistical-mechanical regression and the learning of relationships for nonequilibrium processes.

%\jrgc{Would like say something along the lines of:
%The TIUR is an optimal linear prediction of $\dot{I}$ from $A$.
%The TUR is an optimal linear prediction of $Y$ from $X$.}

\begin{acknowledgments}
This material is based upon work supported by the National Science Foundation under Grant No.~1856250 and the John Templeton Foundation.
J.R.G.\ acknowledges helpful conversations with Bjarne Andresen and Peter Salamon.
\end{acknowledgments}

\vfill

%\bibliography{references}

%apsrev4-2.bst 2019-01-14 (MD) hand-edited version of apsrev4-1.bst
%Control: key (0)
%Control: author (8) initials jnrlst
%Control: editor formatted (1) identically to author
%Control: production of article title (0) allowed
%Control: page (0) single
%Control: year (1) truncated
%Control: production of eprint (0) enabled
%

\end{document}